\documentclass[12pt,a4paper]{article}%
\usepackage{epsfig}
\usepackage{amsmath}
\usepackage{amsfonts}
\usepackage{amssymb}
\usepackage{graphicx}
\usepackage[T1]{fontenc}
\usepackage[utf8]{inputenc}
\usepackage{authblk}%
\providecommand{\U}[1]{\protect\rule{.1in}{.1in}}
\begin{document}

\title{Pseudo-invariant approach for a particle in a complex time-dependent linear potential}
\author{{\small Walid Koussa}$^{a}$\thanks{E-mail: koussawalid@yahoo.com},
{\small Mustapha Maamache}$^{a}$\thanks{E-mail: maamache@univ-setif.dz }\\$^{(a)}${\small Laboratoire de Physique Quantique et Syst\`{e}mes Dynamiques,}\\{\small Facult\'{e} des Sciences, Universit\'{e} Ferhat Abbas S\'{e}tif 1,
S\'{e}tif 19000, Algeria}\textit{.}}
\date{}
\maketitle

\begin{abstract}
The Lewis and Riesenfeld method has been investigated, by Ramos et al in Ref.
\cite{Ram}, for quantum systems governed by time-dependent $\mathcal{PT}$
symmetric Hamiltonians and particularly where the quantum system is a particle
submitted to action of a complex time-dependent linear potential. We discuss
the method they used and propose an alternative one which leads to physically
acceptable uncertainty product and to complex $x$ and $p$ expectation values
but describe the classical motion. We used, for this situation, a linear
pseudo hermitian invariant operator which allow us to solve analytically the
time-dependent Schr\"{o}dinger equation for this problem and to construct a
Gaussian wave packet solution. The normalization condition for the invariant
eigenfunctions with the Dirac delta function is correctly\ obtained, contrary
to what is stated in Ref. \cite{Ram}.



\end{abstract}

\section{Introduction}

One of the "principles" of quantum theory is the association of a Hermitian
operator with any physical quantity, a property that guarantees the reality of
eigenvalues. In reality, the condition of hermiticity is a sufficient
condition, which is by no means necessary, since there are non hermitic
operators whose spectrum is real. The central idea is to replace the condition
of hermiticity by a weaker condition obviously also ensures the reality of
eigenvalues. This led Bender and Boettcher \cite{Bender1998} to propose
replacing the condition of hermiticity by\ the parity-time ($\mathcal{PT}$ )
symmetry, the invariance under simultaneous parity and time reversal
transformation, that plays an important role in non-Hermitian quantum
mechanics, optics physics, condensed matter and quantum field theory. Starting
in quantum mechanics, the concept of $\mathcal{PT}$ symmetry found
applications in many areas of physics \cite{Bend,Rub,Schi}. In particular,
there is a lot of interest in optics due to experimental realizations of
paraxial $\mathcal{PT}$ symmetric optics \cite{Mak,Muss}. Recent applications
include single-mode $\mathcal{PT}$ lasers \cite{52,64} and unidirectional
reflectionless $\mathcal{PT}$ -symmetric metamaterials at optical frequencies
\cite{53}. $\mathcal{PT}$ symmetric systems demonstrate many nontrivial
non-conservative wave interactions and phase transitions, which can be
employed for signal ltering and switching, opening new prospects for active
control of light \cite{133}.

Parity $\mathcal{P}$ has the effect to change the sign of the momentum
operator $p$ and the position operator $x$. The anti-linear operator
$\mathcal{T}$ has the effect to change the sign of the momentum operator $p$
and the pure imaginary complex number $i$. When an eigenstate of the
Hamiltonian is simultaneously an eigenstate of $\mathcal{PT}$ , the
eigenvalues are real we call the symmetry unbroken; otherwise the symmetry is
broken and the eigenvalues come in complex conjugate pairs and violates the
unitarity of the theory. Replacing the standard Hermitian inner product with
the obvious choice
\begin{equation}
\langle\mathcal{\phi}\left\vert \mathcal{\psi}\right\rangle _{\mathcal{PT}%
}=\int_{\mathrm{C}}dx\left[  \mathcal{PT\phi}(x)\right]  \mathcal{\psi}(x)
\label{0}%
\end{equation}
where $\mathcal{PT\phi}(x)=\mathcal{\phi}^{\mathcal{\ast}}(-x)$ and the
integral is taken over the contour in the complex-$x$ plane.The advantage of
this inner product is that the associated norm $\langle\mathcal{\phi
}\left\vert \mathcal{\phi}\right\rangle $ is conserved in time. On unbroken
eigenstates $\left\vert \phi_{n}\right\rangle $ of a $\mathcal{PT}$ -symmetric
Hamiltonian, the inner product (\ref{0}) is (under appropriate assumptions ) pseudo-orthonormal:%

\begin{equation}
\langle\phi_{m}\left\vert \phi_{n}\right\rangle _{_{\mathcal{PT}}}%
=(-)^{m}\delta_{mn} \label{00}%
\end{equation}

Since the $\mathcal{PT}$-norm is not positive-definite, to render the energy
eigenstates orthonormal is to redefine the inner product \ (\ref{0}) by
introducing a new symmetry, denoted $\mathcal{C}$ \cite{Bender2002,Bender2007}%
, having properties very similar to the charge conjugation operator, inherent
in all $\mathcal{PT}$-symmetric Hamiltonians that possess an unbroken
$\mathcal{PT}$ symmetry. This has allowed to introduce an inner-product
structure associated with $\mathcal{CPT}$ conjugation for which the norms of
quantum states are positive definite and unitary-invariant. In particular,
$\mathcal{CPT}$ symmetry is shown to generalize the conventional Hermiticity
requirement by replacing it with a dynamically determined inner product (one
that is defined by the Hamiltonian itself). Several authors have studied time
independent quantum systems governed by non-Hermitian Hamiltonians
\cite{bba,ZA0,Zn,sw,ZA,sw1,ZA1,jda}.

Even before the discovery of $\mathcal{PT}$ -symmetry and the introduction of
the $\mathcal{CPT}$ -inner product, there have been very general
considerations \cite{Scholz} addressing the question of how a consistent
quantum mechanical framework can be constructed from the non-Hermitian
Hamiltonian systems. It was understood at that time that quasi-Hermitian
systems \cite{Scholz} would lead to positive inner products. It has been
clarified \cite{most1,most2,most3,most4} that a non-Hermitian Hamiltonian
having all eigenvalues real is connected to its Hermitian conjugate,
\begin{equation}
H^{\dag}=\eta H\eta^{-1} \label{1}%
\end{equation}
through a linear, Hermitian, invertible and bounded metric operator $\eta
=\rho^{+}\rho$ with a bounded inverse, i.e. $H$ is Hermitian with respect to a
positive definite inner product $\left\langle .,.\right\rangle _{\eta
}=\left\langle .\left\vert \eta\right\vert .\right\rangle $ defined as
\begin{equation}
\langle\phi_{m}^{H}\left\vert \phi_{n}^{H}\right\rangle _{\eta}=\langle
\phi_{m}^{H}|\eta\left\vert \phi_{n}^{H}\right\rangle =\delta_{mn}%
\end{equation}
and called $\eta$ -pseudo-Hermitian inner product. It is also established
\cite{most1,most2,most3,most4} that the non Hermitian Hamiltonian (or a
pseudo-Hermitian Hamiltonian) $H$ can be transformed to an equivalent
Hermitian one given by
\begin{equation}
h=\rho H\rho^{-1}%
\end{equation}
where $h$ is the equivalent Hermitian analog of $H$ with respect to the
standard inner product $\left\langle .,.\right\rangle .$ $\rho$ is often
called the Dyson map \cite{Dys}. Thus, although the eigenvalue spectra of $h$
and $H$ are identical, relations between their eigenvectors will differ
\begin{equation}
\left\vert \psi_{n}^{h}\right\rangle =\rho\left\vert \phi_{n}^{H}%
\right\rangle
\end{equation}

All these efforts have been devoted to study time-independent non-Hermitian
systems. Whereas the treatment for systems with time-dependent non-Hermitian
Hamiltonians with time-independent or time-dependent an metric operators have
been extensively studied
\cite{Faria1,Faria2,most5,znojil2,znojil3,Bila,wang1,wang2,MZ2,
mus,MZ1,fring1,fring2,YG,luiz1,khant,maam,frith,luiz2,mm,MZ,f1,f2,f3,f4,JG1,M1,BB,f5,JG2,JG3,JG4,MAde}%
. Nevertheless, the existence of invariants (constants of the motion or first
integral) introduced by Lewis- Riesenfeld \cite{Lewis} is a factor of central
importance in the study of time-dependent systems.

While research on $\mathcal{PT}$-symmetry has focused on time-independent
Hamiltonians, very few works using a $\mathcal{PT}$-symmetric time-dependent
Hamiltonians, where the time-reversal operator $\mathcal{T}$ has also the
effect to change the sign of the time $t\rightarrow-t$ and whose action on the
wave function defined as \cite{Wi,Ber,chern}
\begin{equation}
\text{ }\mathcal{T}\psi(x,t)=\psi^{\ast}(x,-t) \label{2}%
\end{equation}
is barely found in the literature \cite{ss,y1,y2,NM,luo,luo1,M2}.

In a recent paper, Ramos et al \cite{Ram} extend the well-known Lewis and
Riesenfeld invariant method \cite{Lewis} to $\mathcal{PT}$-symmetric
time-dependent non-Hermitian Hamiltonians and apply it to the quantum motion
of a particle in the presence of a complex time-dependent linear potential
with $\mathcal{PT}$- symmetry. They have misleadingly claim that the invariant
eigenstates normalization condition associated with the Dirac delta function
is verified.

The main objective in this paper is to give, in section 2, a brief recall \ of
results discussed by Ramos et al \cite{Ram} on the motion of a particle under
the action of a complex time dependent $\mathcal{PT}$ -symmetric linear
potential. After that, we discuss the misleadingly results concerning the
$\mathcal{PT}$- inner product of the simultaneously eigenstates of $\ $the
$\mathcal{PT}$ operator and the $\mathcal{PT}$ -symmetric invariant operator
$I^{\mathcal{PT}}(t)$. \ In section 3, we give an alternative method based on
the pseudo-Hermitian invariant operator \cite{maam, M1} to find the solutions
for a particle submitted to the action of a complex time-dependent linear
potential. Finally in section 4, we construct the Gaussian wave packet state
for this problem.\ Despite that the expectation values of the $x$ and $p$
operators are complex, they are identical to the classical variables $x_{c}$ ,
$p_{c}$. In addition, we obtain that the uncertainty product is physically acceptable.

\section{$\mathcal{PT}$ -symmetric invariant operator $I^{\mathcal{PT}}(t)$
for complex time-dependent linear potential \ \ }

B.F.Ramos et al \cite{Ram} have investigated the motion quantum of a particle
with time-dependent mass subject to the action of a complex time-dependent
linear potential described as%
\begin{equation}
H(t)=\frac{p^{2}}{2m(t)}+if(t)x\ \ \ \ \ \ \ \ \ \ \ \label{ham}%
\end{equation}
$\ $\ where $f(t)$ is a real time-dependent function. The classical variables
describing the equations of motion are given by
\begin{equation}
\dot{p}_{c}=-\frac{\partial H}{\partial x_{c}}=-if(t)
\end{equation}
and%

\begin{equation}
\dot{x}_{c}=\frac{\partial H}{\partial p_{c}}=\frac{p_{c}}{m(t)}%
\end{equation}
By solving the above two equations, the space and momentum operators can be
obtained in terms of the initial conditions, given by%
\begin{equation}
p_{c}=p_{0}-i\int_{0}^{t}f(t^{\prime})dt^{\prime}%
\end{equation}
and%

\begin{equation}
x_{c}=x_{0}+p_{0}\int_{0}^{t}\frac{dt^{\prime}}{m(t^{\prime})}-i\left(
\int_{0}^{t}\frac{dt^{\prime}}{m(t^{\prime})}\int_{0}^{t^{\prime}}f(\tau
)d\tau\right)
\end{equation}

By extending the well-known Lewis and Riesenfeld invariant method, they looked
for a $\mathcal{PT}$ symmetric non-Hermitian time-dependent linear operator
given by%
\begin{equation}
I(t)=a(t)x+b(t)p+c(t)
\end{equation}
where $a(t),b(t)$ and $c(t)$ are complex time-dependant c-number functions.

Inserting the invariant $I(t)$ in the Van-Neumann equation%
\begin{equation}
i\frac{\partial I(t)}{\partial t}=\left[  H(t),I(t)\right]  \label{VN}%
\end{equation}
and after some algebra, gives%
\begin{align}
a(t)  &  =a_{0}\nonumber\\
b(t)  &  =-a_{0}\int_{0}^{t}\frac{d\tau}{m(\tau)}\\
c(t)  &  =-ia_{0}\int_{0}^{t}d\tau f(\tau)\int_{0}^{\tau}\frac{d\tau^{\prime}%
}{m(\tau^{\prime})}\nonumber
\end{align}
on the other hand, the$\mathcal{PT}$ symmetric invariant operator condition
\begin{equation}
I^{\mathcal{PT}}(t)=\left(  \mathcal{PT}\right)  I(t)\left(  \mathcal{PT}%
\right)  =I(t) \label{pt}%
\end{equation}
provides
\begin{align}
a_{0}  &  =-a_{0}^{\ast}\nonumber\\
b^{\ast}(t)  &  =-b(t)\label{van}\\
c^{\ast}(t)  &  =c(t)\nonumber
\end{align}
Using the transformation%
\begin{equation}
\varphi_{\lambda}(x,t)\rightarrow\mathcal{\phi}_{\lambda}(x,t)=e^{-i\frac
{\theta_{\lambda}}{2}}\varphi_{\lambda}(x,t)
\end{equation}
and solving the eigenequation%
\begin{equation}
I^{\mathcal{PT}}(t)\mathcal{\phi}_{\lambda}(x,t)=\lambda\mathcal{\phi
}_{\lambda}(x,t)\text{ \ } \label{eigen}%
\end{equation}
so that the eigenfunctions (Eq. (37) of Ref. \cite{Ram}) are given by
\begin{align}
\mathcal{\phi}_{\lambda}(x,t)  &  =\sqrt{\frac{\sigma_{\lambda}}{2\pi\hbar
b(t)}}\exp\left\{  \frac{i}{\hbar b(t)}\left[  \left(  \lambda-c(t)\right)
x-\frac{a_{0}}{2}x^{2}\right]  \right\}  ,\text{\ }\\
\sigma_{\lambda}  &  =\pm1.\nonumber
\end{align}
\ Without loss of generalities, we drop the phase factor $e^{-i\frac
{\theta_{\lambda}}{2}}$. Thus, the eigenstates $\mathcal{\phi}_{\lambda}(x,t)$
of $I^{\mathcal{PT}}(t)$ are eigenstates of the $\mathcal{PT}$ operator \ with
eigenvalue $1,$ when the action of $\mathcal{PT}$ operator on the wave
function is as follows
\begin{align}
\mathcal{PT\phi}_{\lambda}(x,t)  &  =\mathcal{\phi}_{\lambda}^{\ast
}(-x,-t)\nonumber\\
&  =\sqrt{\frac{\sigma_{\lambda}}{2\pi b^{\ast}(-t)}}\exp\left\{  \frac
{-i}{b^{\ast}(-t)}\left[  \left(  \lambda-c^{\ast}(-t)\right)  (-x)-\frac
{a_{0}^{\ast}}{2}x^{2}\right]  \right\} \nonumber\\
&  =\sqrt{\frac{\sigma_{\lambda}}{2\pi b(t)}}\exp\left\{  \frac{i}%
{b(t)}\left[  \left(  \lambda-c(t)\right)  x-\frac{a_{0}}{2}x^{2}\right]
\right\}  =\mathcal{\phi}_{\lambda}(x,t). \label{for}%
\end{align}
On the other hand, the authors of Ref.\cite{Ram} claim, incorrectly, that the
normalization condition (Eq. (39) in \cite{Ram}) associated with the Dirac
delta function is verified. To see that their assertion is not correct, it is
enough to calculate explicitly the $\mathcal{PT}$ inner product defined as%

\begin{align}
\int_{-\infty}^{+\infty}\left[  \phi_{\lambda^{^{\prime}}}(x,t)\right]
^{\mathcal{PT}}\phi_{\lambda}(x,t)dx  &  =\int_{-\infty}^{+\infty
}\mathcal{\phi}_{\lambda^{\prime}}^{\ast}(-x,-t)\text{ }\phi_{\lambda
}(x,t)dx=\int_{-\infty}^{+\infty}\mathcal{\phi}_{\lambda^{\prime}}(x,t)\text{
}\phi_{\lambda}(x,t)dx\nonumber\\
&  =\frac{\sqrt{\sigma_{\lambda^{^{\prime}}}\sigma_{\lambda}}}{2\pi b(t)}%
\int_{-\infty}^{+\infty}\exp\left\{  \frac{i}{b(t)}\left[  \left(
(\lambda+\lambda^{^{\prime}})-2c(t)\right)  x-a_{0}x^{2}\right]  \right\}
dx\neq\delta(\lambda-\lambda^{^{\prime}}). \label{PRO}%
\end{align}
According to the invariant operator theory, the time-dependent Schr\"{o}dinger
equation takes the form
\begin{equation}
\psi_{\lambda}(x,t)=e^{i%
\mu
_{\lambda}(t)}\phi_{\lambda}(x,t),
\end{equation}
where the phase functions are given by%
\begin{equation}%
\mu
_{_{\lambda}}(t)=-\frac{1}{2\hbar}\int_{0}^{t}\frac{\left(  \lambda
-c(\tau)\right)  ^{2}}{m(\tau)b^{2}(\tau)}d\tau.
\end{equation}
So, they construct a Gaussian wave packet solution
\begin{equation}
\Psi(x,t)=\int_{-\infty}^{+\infty}g(\lambda)\psi_{\lambda}(x,t)d\lambda
\label{GS}%
\end{equation}
where the Gaussian weight function $g(\lambda)$ is given by
\begin{equation}
g(\lambda)=\frac{\sqrt{d}}{\sqrt{2\pi}}\exp\left[  -\frac{d^{2}}{4}\lambda
^{2}\right]  .
\end{equation}

Thus, the general solution (\ref{GS}) is given by equation \ (47) of
Ref.\cite{Ram}. Using this wave packet solution, they calculate the
expectation values of the position$\left\langle x\right\rangle _{\mathcal{PT}%
}$ and the momentum $\left\langle p\right\rangle _{\mathcal{PT}}$ as well as
the uncertainty product \ $\Delta x.\Delta p$ , where the expectation value
$\left\langle O\right\rangle _{\mathcal{PT}}$ of an operator $O$ is defined
as
\[
\int_{-\infty}^{\infty}dx\Psi(x,t)^{\mathcal{PT}}\text{ }O\text{ }\Psi(x,t)
\]
They state that the $\mathcal{PT}$ operator acts on the wave function as
follows \cite{Note}
\begin{equation}
\mathcal{PT}\Psi(x,t)=\Psi(x,t)^{\mathcal{PT}}=\Psi^{\ast}(-x,t), \label{c1}%
\end{equation}
which is in contrast with the definition (\ref{for}) employed to show
that\ the eigenstates of the linear invariant are also eigenstates of the
$\mathcal{PT}$ operator.

However, knowing that the wave function is a scalar and taking into account
that $\mathcal{T}$ is antilinear and antiunitary operator, we obtain the time
reversal rule for the wave function $\mathcal{T}\Psi(x,t)=\Psi^{\ast}(x,-t)$.
That is the general rule for time reversal in quantum mechanics: if a
\ certain state is described by the wave function $\Psi(x,t)$, then the
"time-reversed" state is described by the function $\Psi^{\ast}(x,-t)$. The
change to the complex conjugate \ function is necessary because the "correct"
time dependence must be restored, \ after being lost through the change in the
sign of $t$ \ \cite{Wi}.

Finally, they find that the expectation values $\left\langle x\right\rangle
_{\mathcal{PT}}$ , $\left\langle p\right\rangle _{\mathcal{PT}}$ are imaginary
numbers so that the position, momentum operators are not observables. As a
consequence, the uncertainty relation $\Delta x.\Delta p$, which is a complex
number, is physically unacceptable.

In the next section, using the pseudo-Hermitian invariant operator approach
\cite{maam, M1} we get an accepted physical quantities for a " Particle in a
complex time-dependent linear potential ". We show that the expectation values
of $x$ and $p$ are complex numbers that describe the classical motion while
the uncertainty relation is physically acceptable. On the other hand, the
normalization condition \ for \ the invariant eigenfunctions with the Dirac
delta function is verified.

\section{The complex time-dependent linear potential:pseudo-invariant method}

The beginning of this section briefly recalls the results of the
pseudo-invariant operator technique \cite{maam, M1}. In complete analogy to
the time independent scenario a self-adjoint invariant operator $I^{h}(t)$,
i.e., an observable, in the Hermitian system which has an observable
counterpart $I^{PH}\left(  t\right)  $ in the non-Hermitain system are related
to each other as $I^{h}(t)=\rho(t)I^{PH}(t)\rho^{-1}(t)$ $\Leftrightarrow
I^{PH\dag}\left(  t\right)  =\eta(t)I^{PH}\left(  t\right)  \eta^{-1}(t)$ was
introduced and adressed in details in Ref. \cite{maam, M1} that we will
briefly recall. Given a non-Hermitian time-dependent Hamiltonian operator
$H(t)$, it is possible to build a pseudo-invariant operator $I^{PH}(t)$ verifying%

\begin{equation}
\frac{dI^{PH}(t)}{dt}=\frac{\partial I^{PH}(t)}{\partial t}-i\left[
I^{PH}\left(  t\right)  ,H(t)\right]  =0, \label{lewisPH}%
\end{equation}
and obeys the eigenvalue equation:%
\begin{equation}
\text{ \ }I^{PH}\left(  t\right)  \left\vert \phi_{n}^{H}(t)\right\rangle
=\lambda_{n}\left\vert \phi_{n}^{H}(t)\right\rangle , \label{EingPH}%
\end{equation}
where the eigenvalues $\lambda_{n}$ are time-independent and the eigenstates
$\left\vert \phi_{n}^{H}(t)\right\rangle $ of $I^{PH}\left(  t\right)  $ are
orthonormal
\begin{equation}
\left\langle \phi_{m}^{H}(t)\right\vert \eta(t)\left\vert \phi_{n}%
^{H}(t)\right\rangle =\delta_{m,n}.
\end{equation}
The solutions of the Schr\"{o}dinger equation
\begin{equation}
i\hbar\frac{\partial}{\partial t}\left\vert \Phi^{H}(t)\right\rangle
=H(t)\left\vert \Phi^{H}(t)\right\rangle \label{shro}%
\end{equation}
can be written in terms of the eigenfunctions $\left\vert \phi_{n}%
^{H}(t)\right\rangle $ as%
\begin{equation}
\left\vert \Phi_{n}^{H}(t)\right\rangle =e^{i\varphi_{n}(t)}\left\vert
\phi_{n}^{H}(t)\right\rangle ,
\end{equation}
where the phase functions $\varphi_{n}(t)$ are found from the equation:
\begin{equation}
\frac{d\varphi_{n}(t)}{dt}=\left\langle \phi_{n}^{H}(t)\right\vert
\eta(t)\left[  i\frac{\partial}{\partial t}-\frac{H(t)}{\hbar}\right]
\text{\ }\left\vert \phi_{n}^{H}(t)\right\rangle .
\end{equation}

For a particle with time-dependent mass subject to the action of a complex
time-dependent linear potential described by the Hamilonian (\ref{ham}) , we
choose a linear pseudo-Hermitian invariant operator $I^{PH}(t)$ in the form
\begin{equation}
I^{PH}(t)=a(t)\left(  x-\frac{i}{2}\alpha(t)\right)  +b(t)\left(  p-\frac
{i}{2}\beta(t)\right)  +c(t) \label{IPH}%
\end{equation}
where $\alpha(t)$ and $\beta(t)$ are real parameters while $a(t),b(t)$ and
$c(t)\ $are time-dependent \textit{c}-number functions to be determined.

The condition (\ref{lewisPH}) implies that\ \
\begin{align}
\dot{a}(t)  &  =0\nonumber\\
\dot{b}(t)  &  =-\frac{a}{m(t)}\label{VN1}\\
\dot{c}(t)-\frac{i}{2}\left(  \dot{\alpha}a+\dot{\beta}b+\beta\dot{b}\right)
&  =ibf(t)\nonumber
\end{align}
\ after solving these equations, we get%
\begin{align}
a(t)  &  =a_{0}\nonumber\\
b(t)  &  =b_{0}-a_{0}\int_{0}^{t}\frac{1}{m(t^{\prime})}dt^{\prime}%
\label{VN2}\\
c(t)  &  =c_{0}\nonumber\\
f(t)  &  =-\frac{1}{2b}\left(  \dot{\alpha}a_{0}+\dot{\beta}b-\beta\frac
{a_{0}}{m(t)}\right) \nonumber
\end{align}

Since the operator $I^{PH}(t)$ is pseudo-Hermitian, then it fulfills the
condition
\begin{equation}
I^{+PH}(t)=\eta(t)I^{PH}(t)\eta^{-1}(t) \label{quasi}%
\end{equation}
where the operator metric $\eta(t)$ is chosen as%
\begin{equation}
\eta=\rho^{+}\rho=\exp\left[  \beta(t)x-\alpha(t)p\right]  . \label{met2}%
\end{equation}
The condition (\ref{quasi}) provides%
\begin{align}
a(t)  &  =a^{\ast}(t)\nonumber\\
b(t)  &  =b^{\ast}(t)\label{cont}\\
c(t)  &  =c^{\ast}(t)\nonumber
\end{align}
To find a solution of the Schr\"{o}dinger equation of $H(t)$%
\begin{equation}
H(t)\Psi^{H}(x,t)=i\partial_{t}\Psi^{H}(x,t), \label{sch}%
\end{equation}
where $\Psi^{H}(x,t)=\int_{-\infty}^{+\infty}g(\lambda)\psi_{\lambda
}(x,t)d\lambda$ \ \ and \ \ $\psi_{\lambda}(x,t)=e^{i\mu_{\lambda}(t)}%
\varphi_{\lambda}^{I^{PH}}(x,t)$; $\mu_{\lambda}(t)$ should be real, we have
first to solve the eigenvalue equation of the invariant $I^{PH}(t)\varphi
_{\lambda}^{I^{PH}}(x,t)=\lambda\varphi_{\lambda}^{I^{PH}}(x,t)$ . After some
basic calculations, we get that the orthonormalized solutions of the invariant
eigenvalue equation are given by%
\begin{equation}
\varphi_{\lambda}^{I^{PH}}(x,t)=\frac{1}{\sqrt{2\pi b}}\exp\frac{i}{2b}\left[
\left(  2\left(  \lambda-c\right)  +i\beta b\right)  \left(  x-\frac{i}%
{2}\alpha\right)  -a_{0}\left(  x-\frac{i}{2}\alpha\right)  ^{2}\right]
\label{etat}%
\end{equation}
by subtituting $\varphi_{\lambda}^{I^{PH}}(x,t)$ (\ref{etat}) multiplied by a
phase factor $e^{i\mu_{\lambda}(t)}$ in the Schr\"{o}dinger equation
$H(t)\psi_{\lambda}(x,t)=i\partial_{t}\psi_{\lambda}(x,t)$, we get%
\begin{align}
\dot{\mu}_{\lambda}\varphi_{\lambda}^{I^{PH}}(x,t)  &  =\left[  -\frac
{1}{2m(t)b^{2}}\left(  \lambda-c\right)  ^{2}+\frac{1}{2}\left(  \alpha
f-\frac{\dot{\alpha}}{2}\beta+\frac{\beta^{2}}{4m(t)}\right)  \right.
\nonumber\\
&  \left.  -\frac{i}{2b}\left(  \frac{\beta}{m(t)}-\dot{\alpha}\right)
\left(  \lambda-c\right)  \right]  \varphi_{\lambda}^{I^{PH}}(x,t), \label{pp}%
\end{align}
since this phase should be real, it implies that $m\dot{\alpha}=\beta,$ this
is equivalent to $f(t)=-\dot{\beta}(t)/2.$ The phase equation (\ref{pp}) is
simplified into%
\begin{equation}
\dot{\mu}_{\lambda}=\left[  -\frac{1}{2m(t)b^{2}}\left(  \lambda-c\right)
^{2}-\frac{1}{4}\left(  \dot{\beta}\alpha+\frac{\beta^{2}}{2m(t)}\right)
\right]  . \label{phase}%
\end{equation}
So that the general solution can written as%
\begin{equation}
\Psi^{H}(x,t)=\int_{-\infty}^{+\infty}g(\lambda)e^{i\mu_{\lambda}}%
\varphi_{\lambda}^{I^{PH}}(x,t)d\lambda\label{solg}%
\end{equation}
we choose the weight function $g(\lambda)$ in the form
\begin{equation}
g(\lambda)=\sqrt{\frac{\sqrt{d}}{\pi\sqrt{2\pi}b_{0}}}\exp\left[
-d(\lambda-I_{0})^{2}\right]  \exp\left[  -i\frac{d_{0}}{b_{0}}(\lambda
-\frac{I_{0}}{2})\right]  \label{poid}%
\end{equation}
where $\ d,d_{0},I_{0}$ are positive real constants.

After a straighforward calculation, we obtain the general expression solution
in form of the Gaussian wave-packet%
\begin{align}
\Psi^{H}(x,t)  &  =\sqrt{\frac{\sqrt{d}}{\sqrt{2\pi}b\left(  i\int\frac
{1}{2mb^{2}}+d\right)  }}\exp\left\{  -i\int_{0}^{t}\frac{\left(
c-I_{0}\right)  ^{2}}{2mb^{2}}dt^{\prime}\right\} \nonumber\\
&  \exp\left\{  -i\frac{I_{0}}{2}\frac{d_{0}}{b_{0}}\right\}  \exp\left\{
-\frac{i}{4}\int_{0}^{t}\left(  \dot{\beta}\alpha+\frac{\beta^{2}}{2m}\right)
dt^{\prime}\right\} \nonumber\\
&  \exp\left\{  \frac{i}{2b}\left[  (-2\left(  c-I_{0}\right)  +i\beta
b)(x-\frac{i}{2}\alpha)-a_{0}(x-\frac{i}{2}\alpha)^{2}\right]  \right\}
\nonumber\\
&  \exp\left\{  -\frac{\left[  (x-\frac{i}{2}\alpha)-b\left(  -\int_{0}%
^{t}\frac{\left(  c-I_{0}\right)  }{mb^{2}}dt^{\prime}+\frac{d_{0}}{b_{0}%
}\right)  \right]  ^{2}}{4b^{2}\left(  i\int_{0}^{t}\frac{1}{2mb^{2}%
}dt^{\prime}+d\right)  }\right\}  \label{solu}%
\end{align}

Now, we calculate the expectation values of the position and momentum
operators in the Gaussian state $\Psi^{H}(x,t)$%

\begin{align}
\left\langle x\right\rangle _{\eta}  &  =\left\langle \Psi^{H}(t)\right\vert
\eta x\left\vert \Psi^{H}(t)\right\rangle =d_{0}-\frac{c_{0}}{b_{0}}\int
_{0}^{t}\frac{dt^{\prime}}{m(t^{\prime})}+\frac{i}{2}\alpha\label{vmx}\\
\left\langle p\right\rangle _{\eta}  &  =\left\langle \Psi^{H}(t)\right\vert
\eta p\left\vert \Psi^{H}(t)\right\rangle =-\frac{c_{0}}{b_{0}}+\frac{i}%
{2}\beta\label{vmp}%
\end{align}
it is obvious that $\left\langle x\right\rangle _{\eta}$ and $\left\langle
p\right\rangle _{\eta}$ are identical to the classical variables $x_{c}$ ,
$p_{c}$
\begin{equation}
\left\langle x\right\rangle _{\eta}=x_{c}\text{ \ \ \ , \ }\left\langle
p\right\rangle _{\eta}=p_{c} \label{varcl}%
\end{equation}
We also evaluate the uncertainty in the position and the momentum%
\begin{equation}
\Delta x=\sqrt{\left\langle x^{2}\right\rangle _{\eta}-\left(  \left\langle
x\right\rangle _{\eta}\right)  ^{2}}=\frac{b}{\sqrt{d}}\sqrt{d^{2}+\left(
\int_{0}^{t}\frac{dt^{\prime}}{2m(t^{\prime})b^{2}}\right)  ^{2}}%
\end{equation}%
\begin{equation}
\Delta p=\sqrt{\left\langle p^{2}\right\rangle _{\eta}-\left(  \left\langle
p\right\rangle _{\eta}\right)  ^{2}}=\frac{1}{\Delta x}\sqrt{\frac{1}%
{4}+\left[  \frac{1}{4b_{0}db}\int_{0}^{t}\frac{dt^{\prime}}{m(t^{\prime}%
)}-\frac{a_{0}}{b}\left(  \Delta x\right)  ^{2}\right]  ^{2}}%
\end{equation}

as well as the uncertainty product
\begin{equation}
\Delta p\Delta x=\sqrt{\frac{1}{4}+\left[  \frac{1}{4b_{0}db}\int_{0}^{t}%
\frac{dt^{\prime}}{m(t^{\prime})}-\frac{a_{0}}{b}\left(  \Delta x\right)
^{2}\right]  ^{2}}\geqslant\frac{1}{2}, \label{ince}%
\end{equation}
\ \ which is real and greater than (or equal ) to 1/2 and therefore physically acceptable.

\bigskip%

\[%
{\parbox[b]{3.013in}{\begin{center}
\includegraphics[
natheight=2.229500in,
natwidth=2.968900in,
height=2.2693in,
width=3.013in
]%
{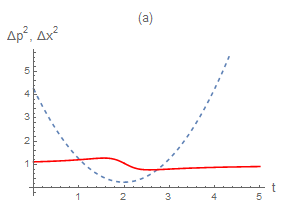}%
\\
Figure1.a.
\ \ \ \ \ \ \ \ \ \ \ \ \ \ \ \ \ \ \ \ \ \ \ \ \ \ \ \ \ \ \ \ \ \ \ \ \ \ \ \ \ \ \ \ \ \ \ \ \ \ \ \ \ \ \ Variances
of the physical position $\left(  \Delta x\right)  ^2$ (Dashed-blue) and
momentum $\left(  \Delta p\right)  ^2$ (solid-red), with the following
parameters:($q_0=p_0=a_0=m=d=1,b_0=2$ and $\hbar=1$).
\end{center}}}
\]

\bigskip

\bigskip%
\[%
{\parbox[b]{3.0441in}{\begin{center}
\includegraphics[
natheight=2.250200in,
natwidth=3.000000in,
height=2.29in,
width=3.0441in
]%
{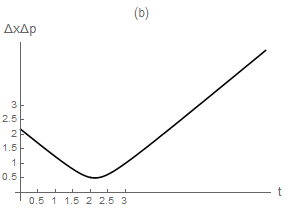}%
\\
Figure1.b.
\ \ \ \ \ \ \ \ \ \ \ \ \ \ \ \ \ \ \ \ \ \ \ \ \ \ \ \ \ \ \ \ \ \ \ \ \ \ \ \ \ \ \ \ \ \ \ \ \ \ \ \ \ \ \ \ \ \ \ \ \ \ \ \ \ \ \ \ \ \ \ \ \ \ \ \ \ \ \ \ \ \ \ \ \ \ \ \ \ \ \ \ The
uncertainty product with the same parameters as in figure (a).
\end{center}}}
\]

\bigskip

\bigskip

The density $\left\vert \rho\Psi^{H}(x,t)\right\vert ^{2}$\ can be written in
function of $\left\langle x\right\rangle _{\eta}$ and $\Delta x$ as
\begin{equation}
\left\vert \rho\Psi^{H}(x,t)\right\vert ^{2}=\left\vert \sqrt{\frac{\sqrt{d}%
}{\sqrt{2\pi}b\left(  i\int_{0}^{t}\frac{dt^{\prime}}{2mb^{2}}+d\right)  }%
}\right\vert ^{2}\exp-\frac{\left[  x-(x_{c}-\frac{i}{2}\alpha)\right]  ^{2}%
}{2\Delta x^{2}}%
\end{equation}
and represents a Gaussian with center in $\left(  \left\langle x\right\rangle
_{\eta}-\frac{i}{2}\alpha\right)  $ $=\left(  x_{0}+p_{0}\int\frac{1}%
{m}\right)  $ and time-dependent width $\Delta x$ and therefore
\begin{equation}
\int_{-\infty}^{+\infty}\left\vert \rho\Psi^{H}(x,t)\right\vert ^{2}dx=1
\label{noR}%
\end{equation}

\[%
{\parbox[b]{3.8821in}{\begin{center}
\includegraphics[
natheight=3.208500in,
natwidth=3.833700in,
height=3.2534in,
width=3.8821in
]%
{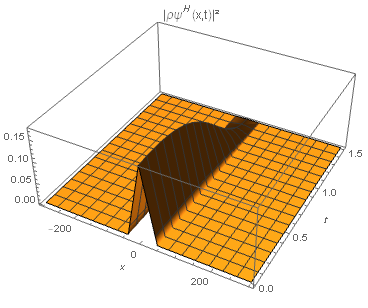}%
\\
Figure 2. Probability density $\left\vert \rho\Psi^H(x,t)\right\vert ^2$for
Gaussian wave packet with $q_0=p_0=a_0=m=d=1,b_0=2$ and $\hbar=1$.The
horizontal and vertical axes correspond to position and time, respectively.
\end{center}}}
\]

In summary, using the time-dependent pseudo-Hermitian linear invariant method,
we have found the solutions for a particle submitted to the action of a
complex time-dependent linear potential. Furthermore, we have constructed a
Gaussian wave packet state for our problem and shown that the time-dependent
probability density associated with this packet is Gaussian and remains
Gaussian for all time.\ In addition, the expected values of the operators $x$
and $p$ , even though that are complex numbers, represent the classical
solutions. We have found that the uncertainty product is physically
acceptable. Also, the normalization condition \ for \ the invariant
eigenfunctions with the Dirac delta function is correctly obtained.


\begin{thebibliography}{99}
\bibitem {Note}The authors of Ref. \cite{Ram} wrote Eq. (\ref{c1}), this means
that eigenstates $\mathcal{\phi}_{\lambda}(x,t)$ of the invariant operator
$I(t)$ are transformed under the $\mathcal{PT}$ action as $\mathcal{PT\phi
}_{\lambda}(x,t)=\mathcal{\phi}_{\lambda}^{\ast}(-x,t).$ This implies that the
eigenstates $\mathcal{\phi}_{\lambda}(x,t)$ of $I(t)$ are \textbf{NOT}
eigenstates of the $\mathcal{PT}$ operator as claimed by the authors of Ref.
\cite{Ram}. One can see that in the following
\begin{align}
PT\phi_{\lambda}(x,t)  &  =\sqrt{\frac{\sigma_{\lambda}}{2\pi b^{\ast}(t)}%
}\exp\left\{  \frac{-i}{b^{\ast}(t)}\left[  \left(  \lambda-\gamma^{\ast
}(t)\right)  (-x)-\frac{a_{0}^{\ast}}{2}x^{2}\right]  \right\} \nonumber\\
&  =\sqrt{\frac{\sigma_{\lambda}}{-2\pi b(t)}}\exp\left\{  \frac{-i}%
{b(t)}\left[  \left(  \lambda-c(t)\right)  x-\frac{a_{0}}{2}x^{2}\right]
\right\}  \text{ \ }\label{ar}\\
&  \neq\phi_{\lambda}(x,t).\nonumber
\end{align}
Morever these eigenstates of $I(t)$ will not also satisfy the normalization
condition with the Dirac delta function
\begin{equation}
\int_{-\infty}^{+\infty}\phi_{\lambda^{^{\prime}}}^{PT}(x,t)\phi_{\lambda
}(x,t)dx=\frac{\sqrt{\sigma_{\lambda^{^{\prime}}}\sigma_{\lambda}}}{2\pi
}\left[  \exp\frac{(\lambda-\lambda^{^{\prime}})}{\left\vert b(t)\right\vert
}x\right]  _{-\infty}^{+\infty} \label{CI}%
\end{equation}


where $b(t)=i\left\vert b(t)\right\vert $ is purely imaginary. \
\end{thebibliography}
\end{document}